\documentclass[a4paper,10pt,twoside]{cpc-hepnp}

\usepackage{multicol}
\usepackage{graphicx}
\usepackage{booktabs}
\usepackage{amssymb,bm,mathrsfs,bbm,amscd}
\usepackage[tbtags]{amsmath}
\usepackage{lastpage}

\usepackage{textcomp}

\begin{document}


\fancyhead[c]{\small F. Liang, \emph{active inductor shunt peaking in high-speed VCSEL driver design}, Submitted to \emph{Chinese Physics C}}
\fancyfoot[C]{\small 2013-0041 Page~\thepage}


\title{
Active inductor shunt peaking in high-speed VCSEL driver design
\thanks{Supported by National Natural Science Foundation of China (11075152) }
}

\author{%
      LIANG Futian$^{1,2}$
\quad GONG Datao$^{2}$
\quad HOU Suen$^{3}$
\quad LIU Chonghan$^{2}$
\quad LIU Tiankuan$^{2}$\\
\quad SU Da-Shung$^{3}$
\quad TENG Ping-Kun$^{3}$
\quad XIANG Annie$^{2}$
\quad YE Jingbo$^{2}$
\quad JIN Ge$^{1,1)}$\email{goldjin@ustc.edu.cn}
}
\maketitle

\address{
$^1$State Key Laboratory of Particle Detection and Electronics, \\University of Science and Technology of China, Hefei Anhui 230026, China\\
$^2$Southern Methodist University, Dallas TX 75275, U.S.A.\\
$^3$Institute of Physics, Academia Sinica, Nangang 11529, Taipei, Taiwan
}

\begin{abstract}
An all transistor active inductor shunt peaking structure has been used in a prototype of 8-Gbps high-speed VCSEL driver which is designed for the optical link in ATLAS liquid Argon calorimeter upgrade. The VCSEL driver is fabricated in a commercial 0.25-$\mu m$ Silicon-on-Sapphire (SoS) CMOS process for radiation tolerant purpose. The all transistor active inductor shunt peaking is used to overcome the bandwidth limitation from the CMOS process. The peaking structure has the same peaking effect as the passive one, but takes a small area, does not need linear resistors and can overcome the process variation by adjust the peaking strength via an external control. The design has been tapped out, and the prototype has been proofed by the preliminary electrical test results and bit error ratio test results. The driver achieves 8-Gbps data rate as simulated with the peaking. We present the all transistor active inductor shunt peaking structure, simulation and test results in this paper.
\end{abstract}





\begin{multicols}{2}
\section{Introduction}

In the High Energy Physics (HEP) experiments, such as the ATLAS, all on-detector electronics systems and devices require the radiation tolerant characteristic. However, the radiation tolerant characteristic is usually out of the concern of commercial products. To meet the requirements, some radiation tolerant commercial products have been proofed and some radiation tolerant ASIC (Application Specific Integrated Circuit) chips have been designed.

VCSEL (Vertical-Cavity Surface-Emitting Lasers) which converts electrical signal into optical is a key component in HEP optical communication link. Its driver amplifies the input signal to meet the requirements of a VCSEL. Nowadays, the commercial VCSEL drivers have a \mbox{10-Gbps} transmission data rate or higher. However, in the HEP applications, the current fastest radiation tolerant VCSEL driver is GBLD~\cite{GBLD}, which is designed to work up to 5-Gbps with a 130~nm CMOS technology.

We designed a prototype of radiation tolerant high-speed VCSEL driver to work at 8-Gbps data rate for the optical link in ATLAS liquid Argon calorimeter upgrade. A commercial 0.25-$\mu m$ Silicon-on-Sapphire (SoS) CMOS process has been used in the design for its radiation tolerant characteristic. The SoS process does not have a transient frequency ($f_{T}$) as fast as the newest process which is commonly used in commercial VCSEL drivers. The bandwidth extension methods must be used to achieve the transmission data rate. Our prototype design proofs that the all transistor active inductor shunt peaking structure can help the driver to achieve the target data rate and overcome the process variations. The technique detail is discussed in section~2. Simulation results are shown in section~3, and on chip preliminary electrical test results are shown in section~4. A summary is in section~5.

\section{Design of active shunt peaking}\label{Design}

The VCSEL driver in our design needs to receive low swing Current-Mode Logic (CML) signals (minimum 2~mA) from line driver in optical link and drive high swing CML signals (up to 8~mA) to VCSEL at 8~Gbps. The targets are required by the optical link. The active inductor shunt peaking is the main method to achieve the data rate with the process we use. For other design details, please refer to our relate work~\cite{LOCLD}.

The shunt peaking techniques are widely used to extend the bandwidth. A passive shunt peaking technique can increase the circuit's bandwidth by nearly 80\% with an inductor in series with a resistor load~\cite{Inductor_peaking}. Even though the on chip inductor is available in our process, it is really inconvenient to use the on chip inductor due to its very large area. Usually, an active inductor, where a transistor combined with a resistor in Fig.~\ref{Fig_Peaking_structure}(a), can be use to overcome the inconvenient of size and accomplish the shunt peaking effect~\cite{APS1}.

Fig.~\ref{Fig_Peaking_structure}(b) is the high frequency small signal model of the conventional active shunt peaking topology in Fig.~\ref{Fig_Peaking_structure}(a). When calculating the equivalent impedance, we apply a test voltage $V_{x}$ at the bottom of the model where is the source of the nMOS M4 in Fig.~\ref{Fig_Peaking_structure}(a), and easily get Eq.~(\ref{eq1}) and (\ref{eq2}). During the calculation, we can cancel the $V_{gs4}$ which voltage crosses the $C_{gs4}$, and get the equivalent impedance $Z_{L}$ as shown in Eq.~(\ref{eq3}).

\begin{eqnarray}
\label{eq1}
\frac{V_{gs4}}{V_{x}}=\frac{\frac{1}{j\omega C_{gs4}}}{R_{2}+\frac{1}{j\omega C_{gs4}}}
\end{eqnarray}
\begin{eqnarray}
\label{eq2}
I_{x}=g_{m4}V_{gs4}
\end{eqnarray}
\begin{eqnarray}
\label{eq3}
Z_{L}=\frac{V_{x}}{I_{x}}=\frac{1}{g_{m4}}+\frac{j\omega C_{gs4}R_{2}}{g_{m4}}
\end{eqnarray}

The first term in Eq.~(\ref{eq3}) is a resistor ($R_{equ}$), and the second term is an inductor ($L_{equ}$). The passive equivalent circuit of the conventional active shunt peaking topology is the same as in the passive inductor shunt peaking technique, shown in Fig.~\ref{Fig_Peaking_structure}(c).

\begin{center}
  \begin{minipage}{8cm}
    \centering
\includegraphics[width=7cm]{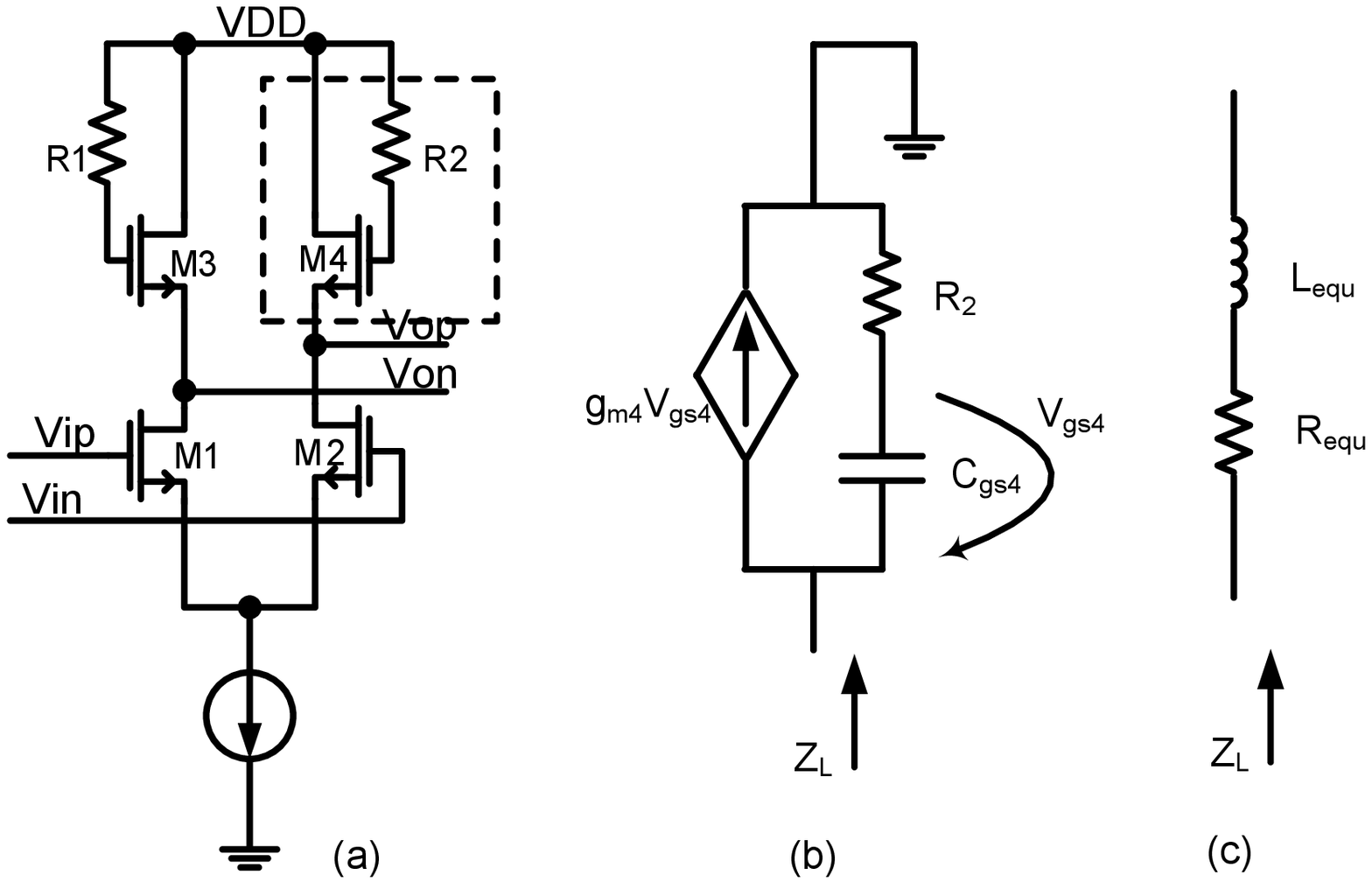}
\figcaption{\label{Fig_Peaking_structure}Conventional active shunt peaking: (a) topology; (b)high frequency small signal model; (c) passive equivalent circuit.}
  \end{minipage}
\end{center}

The equivalent circuit and the capacitive load form a 2nd order RLC network which has a pole in the frequency response. The pole moves the cut-off frequency to a higher range than no peaking case.

From Eq.~(\ref{eq3}), we know the equivalent inductor is determined by $g_{m4}$, $C_{gs4}$ and $R_{2}$. The process variation of a resistor is larger than the other two factors, especially in our SoS process.

To overcome the variation drawback, we use a triode-biased pMOS to replace the resistor shown in Fig.~\ref{Fig_our_APS}. The gates of pMOS M5, M6 have been connected together to an external control pin (Vctrl). By adjusting the voltage applied on gate, we control the equivalent resistance in order to tune the peaking strength.

\begin{center}
  \begin{minipage}{8cm}
    \centering
\includegraphics[height=4.5cm]{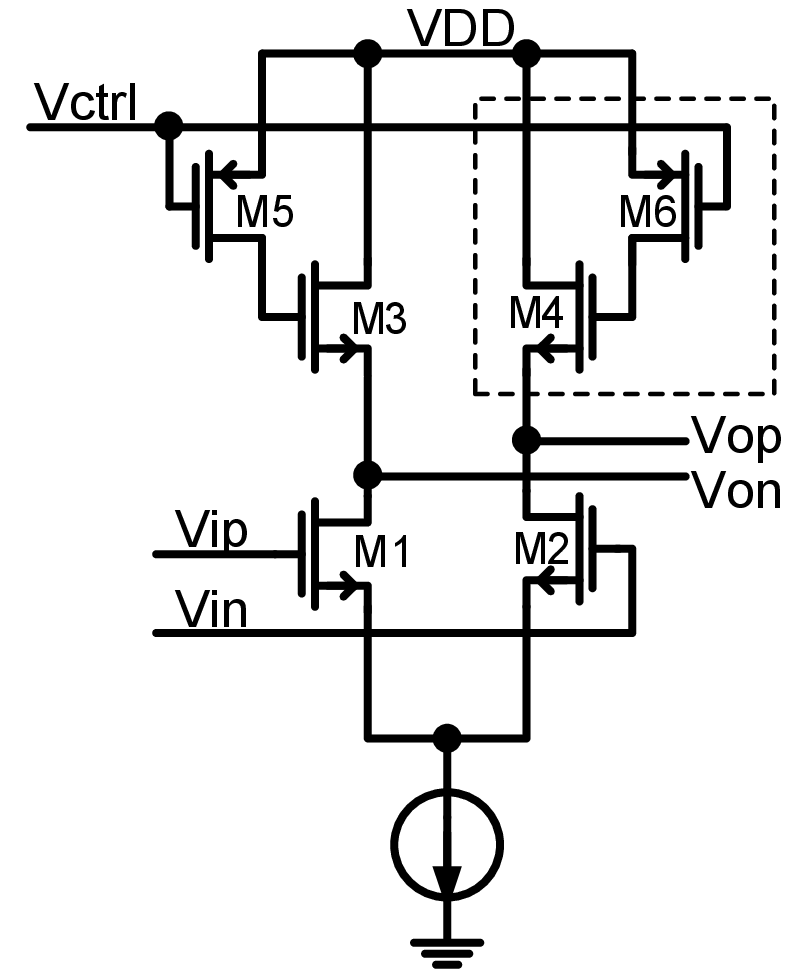}
\figcaption{\label{Fig_our_APS}All transistor active inductor shunt peaking topology}
  \end{minipage}
\end{center}

From the analysis above, we have an all transistor active inductor shunt peaking structure which has the benefit from the passive inductor and only take a small area. The peaking strength is tunable because of the use of a controllable pMOS instead a resistor. The tunable peaking strength make it possible to overcome all the process variations and temperature changes.

\section{Post-layout simulations}\label{Simulations}

We optimized the peaking structure by scanning the parameters of MOSFETs in simulations inside of mathematics. Then we did the post-layout transient simulation of the full VCSEL driver design. In the simulation test bench, the input signal is an 8-Gbps PRBS-7 pattern with 2 mA differential current swing. All the effects from the parameters of the wire-bonding inductor, package capacitor and so on have been taken into our considerations.

Shown in Fig.~\ref{Fig_aps_simulation} is a current eye diagram at the VCSEL load at 27~\textcelsius~typical process corner with default peaking strength from the post-layout simulation. The Deterministic jitter ($D_{j}$) is 4.124~ps, and the eye vertical opening is 7.751~mA (775.1~mV on 100~$\Omega$ load). The output is fine to drive a physical VCSEL.

\begin{center}
  \begin{minipage}{8cm}
    \centering
\includegraphics[width=8cm]{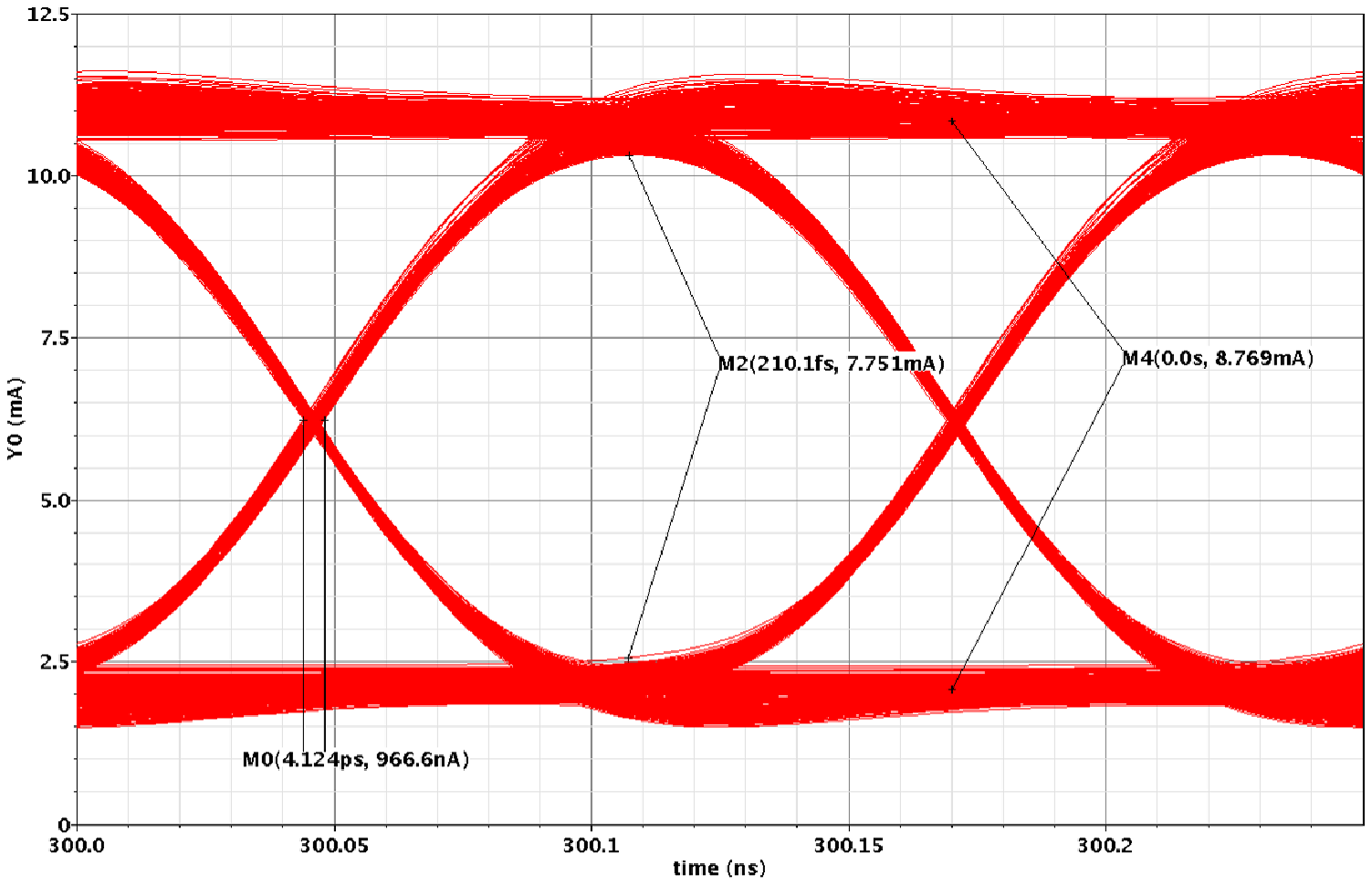}
\figcaption{\label{Fig_aps_simulation}An eye diagram of current passing the VCSEL load at 27~\textcelsius\ typical process corner}
  \end{minipage}
\end{center}

The peaking strength can be adjusted by the Vctrl pin, and a series of simulations have been preformed to insure the tunable peaking strength could cover all the process variations and temperature changes. We run simulation scripts to scan all nine pre-defined process corners and three typical temperatures (27~\textcelsius, 55~\textcelsius, 85~\textcelsius).

Because of the type of MOSFET we use in the peaking structure, the pre-defined process corner --- slowNfastPslowO is the worst one in all simulations. High temperature also effect the circuit functions most.  We use the simulation results of 85~\textcelsius~slowNfastPslowO process corner to demo the coverage of the tunable peaking strength. The tendencies of the deterministic jitter ($D_{j}$) and the vertical eye opening of the output current signal at different peaking control voltages (Vctrl) are shown in Fig.~\ref{Fig_Peaking_simulation}.

\begin{center}
  \begin{minipage}{8cm}
    \centering
    \includegraphics[width=7.8cm]{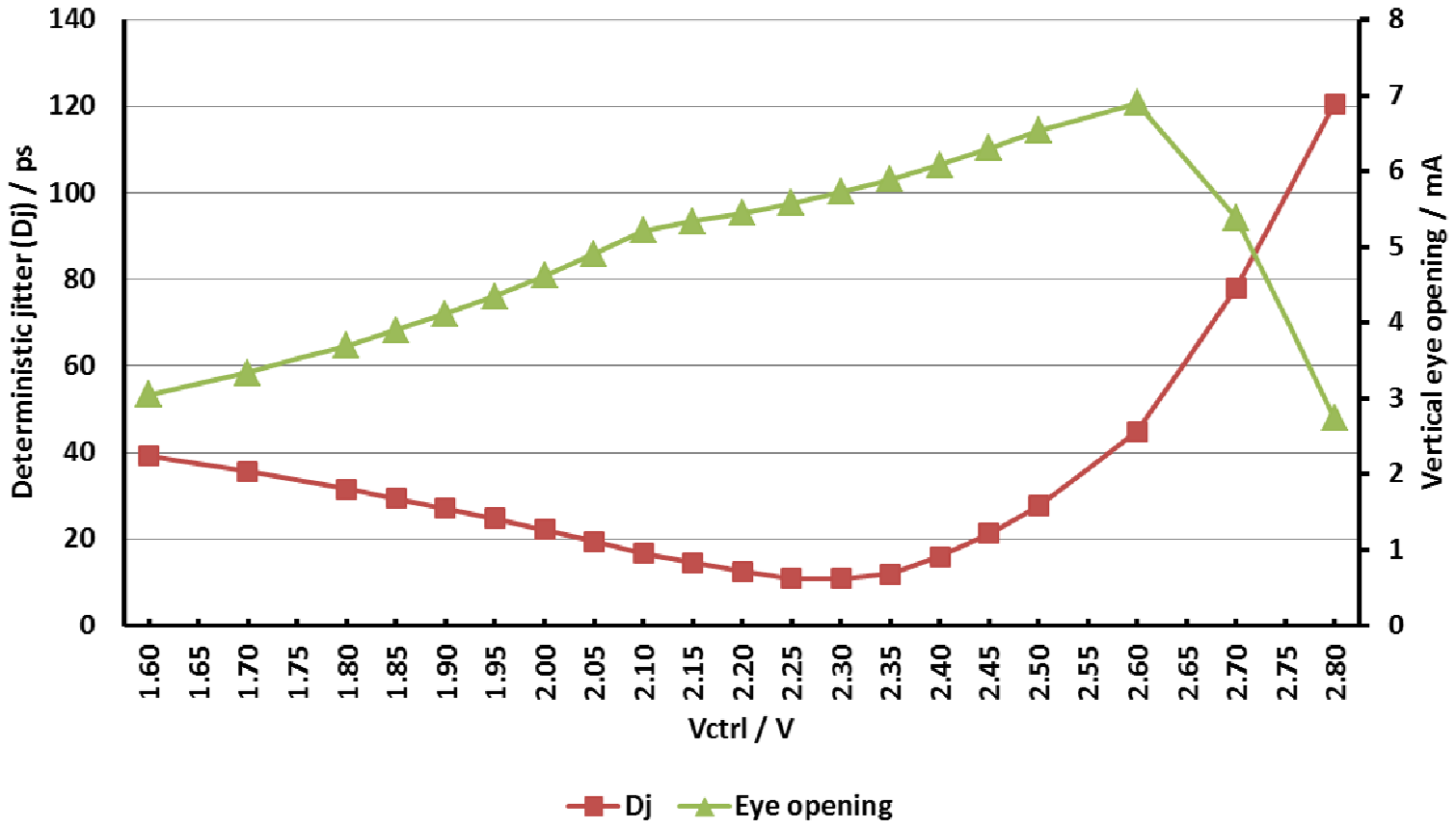}
    \figcaption{\label{Fig_Peaking_simulation}Peaking control tendencies in 85~\textcelsius\ slowNfastPslowO process corner}
  \end{minipage}
\end{center}

In all the process corners and temperatures, not only the 85~\textcelsius~slowNfastPslowO process corner, the simulation results have the similar tendencies. In each tendency curve, we can find a best Vctrl to balance the jitter and eye opening. The active shunt peaking function works as we expected.

\section{On chip tests and results}\label{Test results}

Some quick tests have been done when the chip was fabricated in a Multi Project Wafer (MPW) run.

The test bench of an electrical test is shown in Fig.~\ref{Fig_Test_bench}. The input signal is a 200~mV differential 8-Gbps PRBS-7 pattern which is provided by a pattern generator and attenuators. The input is AC coupled to the test PCB and the output is AC coupled to a sampling oscilloscope. The VCSEL driver has all the default settings except the peaking control.

\begin{center}
  \begin{minipage}{8cm}
    \centering
\includegraphics[width=8cm]{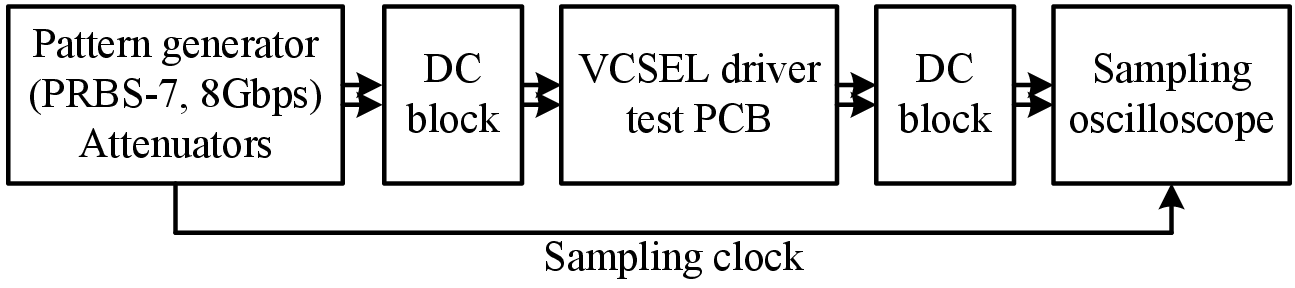}
\figcaption{\label{Fig_Test_bench}Test bench of an electrical test}
  \end{minipage}
\end{center}

The eye diagram of the electrical output and jitter analysis from the sampling oscilloscope are shown in Fig.~\ref{Fig_Testresults_eye}. When the chip was taped out, we didn't know the exact process variation for each chip, and we cannot compare the test results with any existing simulation results directly. 

\end{multicols}
\ruleup
\begin{center}
  \begin{minipage}{\textwidth}
    \centering
\includegraphics[width=.7\textwidth]{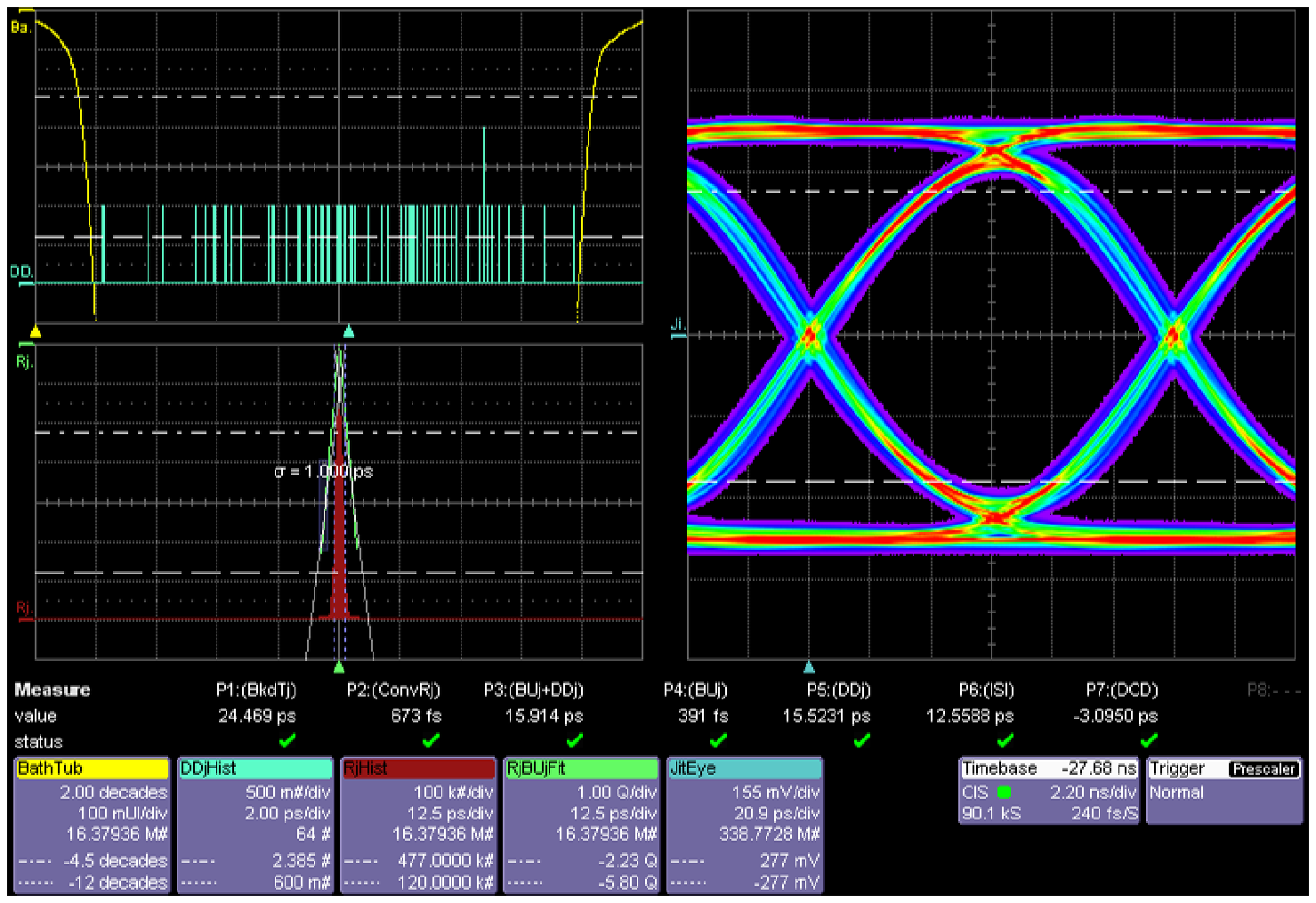}
\figcaption{\label{Fig_Testresults_eye}Eye diagram and jitter analysis in an electrical test}
  \end{minipage}
\end{center}

\ruledown
\begin{multicols}{2}

The total jitter of eye diagram is about 24.469~ps and the eye vertical opening is about 600~mV on 100~$\Omega$ load. These test results close to our simulation results.

Different external peaking control voltages are used in the test, and the tendency is close to the simulation.

We finished a bit error ratio test (BERT) with our driver driving a commercial 10-Gbps VCSEL. The optical fiber and receiver used in the test are all commercial products which have a data rate higher than 8~Gbps. So if an error occurs during the test, it may be mainly caused by the driver. The test results are shown in Table~\ref{BERT}.

\begin{center}
\tabcaption{ \label{BERT}  Results of optical bit error ratio tests}
\footnotesize
\begin{tabular*}{80mm}{rrccr}
\toprule
Data rate & Test time   & Total bits  & Errors & Error ratio \\
\midrule
5.0 Gbps & 300 s & $1.50\times10^{12}$ & 0& $<6.67\times10^{-13}$ \\
8.0 Gbps & 300 s & $2.40\times10^{12}$ & 0& $<4.17\times10^{-13}$ \\
10.0 Gbps & 300 s & $3.00\times10^{12}$ & 0& $<3.33\times10^{-13}$ \\
10.0 Gbps & 54000 s & $5.40\times10^{14}$ & 6& $1.11\times10^{-14}$\\
10.5 Gbps & 300 s & $3.15\times10^{12}$ & 3& $9.52\times10^{-13}$\\
\bottomrule
\end{tabular*}
\end{center}

The target data rate of our design is 8~Gbps, and a bit error ratio is required being smaller than $10^{-12}$ from the industrial standard. From the test results, the driver can work at 10~Gbps with a bit error ratio smaller than the industrial standard in a long time test.

As a prototype, the design is successful from the above test results. The next version with digital configuration has been prepared base on this prototype.

\section{Conclusion}\label{Conclusion}

With the use of all transistor active inductor shunt peaking structure, we successfully designed a prototype of an 8-Gbps VCSEL driver with a commercial 0.25-$\mu m$ SoS CMOS process for radiation tolerant purpose in the optical link in ATLAS liquid Argon calorimeter upgrade. It is proofed that the peaking structure can overcome the bandwidth limitation and process variations. The peaking function works as we expected from the preliminary electrical test results. The bit error ratio test passes the industrial standard. The prototype provides a good reference for the next version of our radiation tolerant VCSEL driver design.
\\

\acknowledgments{The authors acknowledge the department of physics in Southern Methodist University for offering the opportunity to doing the test work.}

\end{multicols}

\vspace{-1mm}
\centerline{\rule{80mm}{0.1pt}}
\vspace{2mm}

\begin{multicols}{2}

\end{multicols}

\clearpage

\end{document}